\documentclass[12pt,a4paper,final]{iopart}

\usepackage{iopams}  
\usepackage{cite,float}
\usepackage{graphicx}
\usepackage[breaklinks=true,colorlinks=true,linkcolor=blue,urlcolor=blue,citecolor=blue]{hyperref}

\begin{document}

\title[Analytic solution of the resolvent equations]{Analytic solution of the resolvent equations for heterogeneous random graphs: spectral and localization properties}

\author{Jeferson D Silva$^{1}$ and Fernando L Metz$^{1,2,*}$}
\address{$^1$ Physics Institute, Federal University of Rio Grande do Sul, 91501-970 Porto Alegre, Brazil.}
\address{$^2$ London Mathematical Laboratory, 8 Margravine Gardens, London W6 8RH, United Kingdom.}
\address{$^*$ Author to whom any correspondence should be addressed.}
\ead{fmetzfmetz@gmail.com}

\begin{abstract}
The spectral and localization properties of heterogeneous random graphs are determined by the resolvent distributional
equations, which have so far resisted an analytic treatment.
We solve analytically the resolvent equations of random graphs with an arbitrary degree distribution in the high-connectivity
limit, from which we perform a thorough analysis of the impact of degree fluctuations on
the spectral density, the inverse 
participation ratio, and the distribution of the local density of states. We show
that all eigenvectors are extended and that the spectral density exhibits a logarithmic or a power-law divergence when the variance of the degree distribution is large enough.
We elucidate this singular behaviour by showing that the distribution of the local density of states at the center of the spectrum
displays a power-law tail determined by the variance of the degree distribution.
In the regime of weak degree fluctuations the spectral density has a finite support, which promotes the stability
of large complex systems on random graphs. 
\end{abstract}

\section{Introduction}

The adjacency matrix of random graphs stores the interactions between the constituents
of large complex systems \cite{newman2018book} ranging from physical and biological to social and technological systems.
The empirical spectral density of the adjacency  matrix and
the localization of its eigenvectors are important to understand
algorithms for node centrality \cite{Restrepo2006,Travis2014} and community detection \cite{Luxburg2007,Nadakudity2012}, as well as
the interplay between the structure of networks and dynamical processes on them. In fact, the
leading eigenpair of the adjacency matrix governs the spreading of diseases \cite{Goltsev2012,Silva2021}, the synchronization
transition \cite{Restrepo2005,Rodrigues2016}, and the linear stability of large complex systems \cite{May1972,Sompolinsky1988,Suweis2015,Neri2020}.
In condensed matter physics, models defined on
random graphs represent mean-field versions
of finite-dimensional lattices which mimic the effects of finite coordination number.
The spin-glass transition and Anderson localization have been intensively investigated on random
graph structures over the past years \cite{Abou1973,Mezard2001}, and they continue to attract
a lot of interest \cite{Lupo2019,Metz2022,Tarzia2022,Colmenarez2022,Biroli2022}. 

The spectral and localization properties of the random adjacency matrix
are determined by the resolvent matrix $\boldsymbol{G}$. The average of its diagonal elements $G_{ii}$
yields the empirical spectral density, while the average of $|G_{ii}|^2$ gives the inverse participation ratio (IPR) \cite{Fyodorov91,Metz2010,Tapias2022}, which
characterizes the volume of the eigenvectors. The full probability density of $G_{ii}$ fulfills
a system of distributional equations, derived in a series of fundamental works \cite{Dean2002,Rogers2008,Kuhn2008} using
the cavity and the replica methods of spin-glass theory (see \cite{Susca2021} for a review of these techniques). The resolvent distributional equations
are exact on locally tree-like random graphs \cite{Bordenave2010} and they provide a solid framework to investigate the spectral
properties of {\it sparse} and {\it heterogeneous} random graphs.
Heterogeneity broadly refers to local fluctuations in the graph structure, such as randomness in the degrees or in the interaction strengths between the nodes, while sparseness means
that the average degree is finite (the degree counts the number of edges attached to a node).
The numerical solutions of the resolvent equations have led to a profusion of results
for the spectral and localization properties of random graphs \cite{Metz2010,Biroli2010,Slanina2012} with
different topological features, including short loops \cite{Metz2011}, modularity \cite{Reimer2011}, and degree-degree correlations \cite{Rogers2010}.
Analogous resolvent equations describe the spectral properties of stochastic matrices on random graphs \cite{Reimer2015,Tapias2022}.

Despite the resolvent distributional equations have led to a tremendous progress in the field, they admit analytic solutions only for 
sparse regular graphs \cite{Bordenave2010,Metz2011}, whose local structure is homogeneous, and
for high-connectivity random graphs \cite{Rogers2008}, where the mean degree is infinitely large and the graph becomes
homogeneous on account of the law of large numbers. In each case, the distribution
of $G_{ii}$ is a Dirac-$\delta$ and the spectral density follows from a simple algebraic equation for the average resolvent.
In a recent paper \cite{Metz2020}, the resolvent equations for
the configuration model of random graphs with a geometric degree distribution have been studied in the high-connectivity limit \cite{Molloy1995,newman2018book,Fosdick2018}.
In this case, the average resolvent
fulfills a transcendental equation and the spectral density diverges at center of the spectrum \cite{Metz2020}.
These analytic findings are interesting for at least two reasons. First, they imply that the spectral density
of high-connectivity random graphs is not generally given by the Wigner law of random matrix theory \cite{Livan2018}, but it explicitly depends
on the degree distribution. Indeed, as rigorously proven in \cite{Dembo2020}, the Wigner universality only holds for degree distributions
that become sharply peaked in the high-connectivity limit.
Second, these findings hint the existence of a rich and nontrivial family of analytic
solutions of the resolvent equations,
sandwiched between the sparse and the dense regime, in which the average degree is
large but the network heterogeneities are still relevant for the spectral properties. The analytic results in \cite{Metz2020} are limited, however, to a geometric
degree distribution.

In this paper we generalize the results of \cite{Metz2020} and we extract the analytic solution of the resolvent
equations for random graphs with arbitrary degree distributions in the high-connectivity limit. To our knowledge, this is the first
example of a full analytic solution for the probability density of  $G_{ii}$  for undirected
random graphs with an heterogeneous structure. We show that the spectral density, the inverse participation ratio, and the distribution
of the local density of states 
are fully determined by the choice of the degree distribution.
We present explicit results for a negative binomial degree distribution, in which the variance 
of the degrees is controlled by a single parameter $0 < \alpha < \infty $ that enables to interpolate between
homogeneous ($\alpha \rightarrow \infty$) and
strongly heterogeneous graphs ($\alpha \rightarrow 0$). In this way, we are able to thoroughly investigate the
impact of degree fluctuations on the spectral and localization properties.
We show that the spectral density undergoes a transition as
the degree fluctuations become stronger. For $\alpha > 1$, the spectral density is a regular function, whereas it
displays either a power-law or a logarithmic divergence at the zero eigenvalue provided $\alpha \in (0,1)$ or $\alpha=1$, respectively.
In the regime of weak degree fluctuations ($1 \ll \alpha < \infty$), the spectral density has a finite support and
large complex systems interacting through the underlying adjacency matrix can be found in a (linearly) stable state.
From the analytic results for the inverse participation ratio and the distribution of the local density
of states, we show that all eigenvectors of the adjacency matrix are extended for any  $\alpha$.
In particular, the distribution of the local density of states at the zero eigenvalue exhibits a power-law
tail with exponent $\alpha+1$, which emphasizes the prominent role of the degree fluctuations and clarifies
the singular behaviour of the spectral density. 
We support our theoretical findings by comparing them with numerical diagonalizations
of large adjacency matrices and, as a byproduct, we show that the adjacency matrix
of a single graph instance can be decomposed for large average degree as a product between a Gaussian random matrix \cite{Livan2018} and the square root of the degree matrix. Such decomposition
provides a straightforward way to sample the adjacency matrix of the configuration model of networks in the
high-connectivity limit.

The paper is organized as follows. In the next section we introduce the random graph model and the resolvent equations
for its adjacency matrix. In section \ref{sec2} we derive the distributional equations and the analytic
expression for the probability density of the resolvent in the high-connectivity limit using the law of large numbers. In section \ref{sec3} we discuss
explicit results for the spectral density, the inverse participation ratio, and the
distribution of the local density of states in the case of a negative binomial degree distribution.
We present a summary and a discussion of our results in section \ref{sec4}, and we provide a more rigorous derivation of the analytic solution of the resolvent distributional equations
in the appendix.


\section{The general setting}
\label{sec1}

We consider a simple and undirected random graph with $N$ nodes. The graph structure is specified by the set of
binary random variables $\{c_{ij} \}$ ($i,j=1,\dots,N$), in which $c_{ij}=c_{ji}=1$ if
there is an undirected edge between nodes $i$ and $j$ ($i \neq j$), and $c_{ij}=0$ otherwise. In addition, we associate
a symmetric coupling strength  $J_{ij} = J_{ji} \in \mathbb R$ to each edge $i\leftrightarrow j$. The degree $k_i = \sum_{j=1}^N c_{ij}$
of a node $i$ gives the number of nodes attached to $i$, and the degree distribution
\begin{equation}
\label{pk_dist}
    p_k = \lim_{N \rightarrow \infty} \frac{1}{N}\sum_{i=1}^N\delta_{k,k_i}
\end{equation}
gives the fraction of nodes with degree $k$. The average degree stands as
\begin{equation}
\label{c_def}
    c = \sum_{k=0}^\infty k  p_k.
\end{equation}

We study the spectral properties of the $N\times N$ adjacency random matrix $\boldsymbol{A}$, with elements
\begin{equation}
\label{Aij_def}
    A_{ij} = c_{ij}J_{ij},
\end{equation}
where the coupling strengths $J_{ij}$ are, apart from the symmetry constraint $J_{ij}=J_{ji}$, independently and identically distributed random variables drawn from a distribution
$p_J$ with mean $J_0/c$ and standard deviation $J_1/\sqrt{c}$. The nonzero values of $\{c_{ij} \}$ are randomly assigned following the 
configuration model of networks \cite{Molloy_etc1,Newman1,Fosdick_2018}, in which a single graph instance is uniformly chosen at random
from the set of all random graphs with a given degree sequence $k_1,\ldots,k_N$ sampled from $p_k$. The configuration model
allows us to fix the degree distribution from the outset and study its impact on the spectral properties.

The adjacency matrix $\boldsymbol{A}$ has a complete set $\{\boldsymbol{v}_\mu\}_{\mu=1,\ldots,N}$ of orthonormal eigenvectors
that fulfill
\begin{equation}
\boldsymbol{A} \boldsymbol{v}_\mu = \lambda_\mu \boldsymbol{v}_\mu,
 \end{equation}
with $\{\lambda_\mu\}_{\mu=1,\ldots,N}$ the set of eigenvalues. The empirical spectral density of $\boldsymbol{A}$ reads
\begin{equation}
\label{rho_A1}
    \rho(\lambda) = \lim_{N \rightarrow \infty} \frac{1}{N}\sum_{\mu=1}^N \delta(\lambda - \lambda_\mu).
\end{equation}
The inverse participation ratio (IPR) of an eigenvector $\boldsymbol{v}_\mu$ is defined as
\begin{equation}
\label{Y_mu}
    Y_{\mu} = \sum_{i=1}^N (v_{\mu,i})^4,
\end{equation}
where $v_{\mu,i}$ is the $i$-th component of $\boldsymbol{v}_\mu$.
The IPR distinguishes between localized and extended eigenvectors in the large $N$ limit.
The components of a localized eigenvector are nonzero on a finite number of nodes and the corresponding
IPR is of order $\mathcal{O}(N^0)$, whereas an extended eigenvector is spread over a finite fraction of nodes
and the IPR vanishes as $\mathcal{O}(1/N)$ in the large $N$ limit.
Since the extent of the eigenvectors typically depends on the corresponding eigenvalue, it is sensible to introduce the eigenvalue-dependent
IPR \cite{Fyodorov91,Mirlin2000,Metz2010,Tapias2022}
\begin{equation}
\label{P_A1}
    \mathcal{I}(\lambda) = \lim_{N \rightarrow \infty} \frac{  \sum_{\mu=1}^N\delta(\lambda-\lambda_\mu)Y_\mu}{  \sum_{\mu=1}^N\delta(\lambda-\lambda_\mu)},
\end{equation}
which is the average of $Y_{\mu}$ over all eigenvectors in an infinitesimal
spectral window around $\lambda$.

The spectral properties of $\boldsymbol{A}$ follow from the resolvent matrix
\begin{equation}
\label{G_def}
    \boldsymbol{G}(z) = (\boldsymbol{I} z-\boldsymbol{A})^{-1},
\end{equation}
where $\boldsymbol{I}$ is the $N \times N$ identity matrix and $z=\lambda- i\epsilon$ lies in the lower complex half-plane.
The  diagonal elements of $\boldsymbol{G}$ determine the spectral density and the eigenvalue-dependent IPR
for a finite regularizer $\epsilon > 0$ according to \cite{Mirlin2000}
\begin{eqnarray}
\label{rho_A2}
   & \rho_{\epsilon}(\lambda)= \frac{1}{\pi}  \lim_{N \rightarrow \infty} \frac{1}{N} \sum_{i=1}^N \mathrm{Im}G_{ii}(z),\\
   \label{P_A2}
   &\mathcal{I}_{\epsilon}(\lambda) = \frac{\epsilon}{\pi\rho_{\epsilon}(\lambda)} \lim_{N \rightarrow \infty} \frac{1}{N} \sum_{i=1}^N |G_{ii}(z)|^2.
\end{eqnarray}
Working with finite $\epsilon$ amounts to replace the Dirac-$\delta$
distributions appearing in Eqs. (\ref{rho_A1}) and (\ref{P_A1}) by Cauchy distributions with a scale parameter $\epsilon$ \cite{Kuhn2008,Rogers2008}.
The spectral observables $\rho(\lambda)$ and $\mathcal{I}(\lambda)$ are reconstructed by taking the
limit $\epsilon \rightarrow 0^{+}$ in Eqs. (\ref{rho_A2}) and (\ref{P_A2}).

By introducing the joint probability density of the real and imaginary parts of $G_{ii}(z)$,
\begin{equation}
\label{P_G1}
    \mathcal{P}_{z}(g) = \lim_{N\to \infty}\frac{1}{N}\sum_{i=1}^N \delta\left[g- G_{ii}(z)\right],
\end{equation}
the spectral density and the IPR can be written as
\begin{eqnarray}
\label{rho_1}
   & \rho_{\epsilon}(\lambda)= \frac{1}{\pi} \mathrm{Im}\langle G \rangle_{\mathcal{P} },\\
\label{P_1}
   &\mathcal{I}_{\epsilon}(\lambda) = \frac{\epsilon}{\pi\rho_{\epsilon}(\lambda)}\langle |G|^2\rangle_{\mathcal{P} },
\end{eqnarray}
where we defined the expectation value
\begin{equation}
  \langle f(G) \rangle_{\mathcal{P} } = \int_{\mathbb H^+} dg  \,f(g) \mathcal P_z(g)
  \label{expect}
\end{equation}
of an arbitrary function $f(g)$ of the random variable $g \in \mathbb{C}$ distributed according to $\mathcal{P}_z(g)$. The
symbol $\mathbb{H}^+$ represents the complex upper half-plane and we have introduced the shorthand notation $dg = d {\rm Re} g \, d {\rm Im} g$.

We see that $\rho_{\epsilon}(\lambda)$ and $\mathcal{I}_{\epsilon}(\lambda)$ are determined by the moments of $\mathcal{P}_z(g)$. In the
limit $N \rightarrow \infty$, the local structure around a randomly chosen node of a graph drawn from the configuration model
converges to a tree \cite{Bordenave2010}, and the probability of finding a short loop in a finite neighbourhood of the
node in question goes to zero. This property implies that the resolvent diagonal elements for a single
graph instance fulfill the equations \cite{Rogers2010}

\begin{equation}
    \label{resolven_eq1}
    G_{ii}(z) = \frac{1}{z - \sum_{j\in \partial_i}J_{ij}^2 G_{jj}^{(i)}(z)} \quad (i=1,\dots,N),
\end{equation}
where $\partial_i$ is the set of nodes adjacent to $i$. The complex variable $G_{jj}^{(i)}$ is the $j$th-diagonal
element of the resolvent on a graph in which node $i \in \partial_j$ and all its incident edges have been deleted \cite{Metz2019}. The variables $\{ G_{jj}^{(i)} \}$ are determined from
the fixed-point solutions of the so-called cavity equations
\begin{equation}
    \label{resolven_eq_cav1}
    G_{jj}^{(i)}(z) = \frac{1}{z - \sum_{\ell \in \partial_j\backslash i}J_{j\ell}^2 G_{\ell\ell}^{(j)}(z)  }  \quad i \in \partial_j,
\end{equation}
where $\partial_j \backslash i$ is the set of nodes connected to $j$ excluding $i$. The total number of cavity
variables $\{ G_{jj}^{(i)} \}$, defined on the edges of the graph, is $\sum_{i=1}^N k_i$. The solutions of Eq. (\ref{resolven_eq_cav1}) lead
to approximations for the resolvent diagonal elements of single graph instances when $N$ is large. In the
limit $N \rightarrow \infty$, Eqs. (\ref{resolven_eq1}) and (\ref{resolven_eq_cav1}) become exact and it is more convenient to
work with the distributions of $G_{ii}(z)$ and $G_{jj}^{(i)}(z)$. Given that both
sides of Eq. (\ref{resolven_eq1}) are equal in distribution, the probability density $\mathcal{P}_z(g)$ is determined from
\begin{equation}
  \label{pdfresolven}
  \fl
    \mathcal{P}_{z}(g) = \sum_{k=0}^\infty p_k \int_{\mathbb{H^+}} \bigg[ \prod_{\ell=1}^{k}d g_\ell  \mathcal{Q}_{z}(g_\ell) \bigg]
\int_{\mathbb{R}} \bigg[\prod_{\ell=1}^{k} d {J_{\ell}} p_J(J_\ell)\bigg]
    \delta\Bigg (g - \frac{1}{z - \sum_{\ell=1}^{k}J_{\ell}^2 g_\ell}\Bigg), 
\end{equation}
where $\mathcal{Q}_{z}(g)$ is the probability density of the cavity variables $G_{jj}^{(i)}(z)$, defined
as
\begin{equation}
  \label{judadd}
\mathcal{Q}_{z}(g) = \lim_{N \rightarrow \infty} \frac{\sum_{i,j=1}^N c_{ij}  \delta\left[g- G_{ii}^{(j)}(z)\right]   }{\sum_{i,j=1}^N c_{ij} },
\end{equation}  
which solves the self-consistent equation
\begin{equation}
  \label{pdfresolvencav_Ninf}
  \fl
\mathcal{Q}_z(g) = \sum_{k=1}^\infty \frac{k}{c}p_k \int_{\mathbb{H}^{+}} \bigg[\prod_{\ell=1}^{k-1}d g_\ell  \mathcal{Q}_z(g_\ell)  \bigg]
 \int_{\mathbb{R}} \bigg[\prod_{\ell=1}^{k-1} d {J_{\ell}} p_J(J_\ell)\bigg]
\delta\Bigg (g - \frac{1}{z - \sum_{\ell=1}^{k-1}J_{\ell}^2 g_\ell}\Bigg).
\end{equation}
Equations (\ref{pdfresolven}) and (\ref{pdfresolvencav_Ninf}) are the distributional version of the resolvent equations. Once we solve
Eq. (\ref{pdfresolvencav_Ninf}) and find a fixed-point solution for $\mathcal{Q}_z(g)$,
the probability density
$\mathcal{P}_{z}(g)$ of the diagonal elements of the resolvent follows from Eq. (\ref{pdfresolven}).
As we will consider the high-connectivity limit $c \rightarrow \infty$, it is interesting to introduce the
joint probability density $\mathcal{W}_{z}(s)$ of the complex variable
\begin{equation}
  \label{S}
S(z) \stackrel{d}{=}  \sum_{\ell=1}^{k}J_{\ell}^2 g_\ell,
\end{equation}
which consists of a sum of independent and identically distributed random variables.
The distribution $\mathcal{P}_{z}(g)$ is written in terms of $\mathcal{W}_{z}(s)$ as
\begin{equation}
\mathcal{P}_{z}(g) = \int_{\mathbb{H}^{+}} d s \mathcal{W}_{z}(s)  \delta\Bigg (g - \frac{1}{z - s}\Bigg).
\end{equation}  
In the context of tight-binding models for the diffusion of an electron on a graph \cite{Abou1973,Parisi2019}, $S(z)$ is known as the self-energy and its
distribution $\mathcal{W}_{z}(s)$ plays an important role in the study of the Anderson localization transition. The average of
a function of $G_{ii}(z)$, defined by Eq. (\ref{expect}), can be recast in terms $\mathcal{W}_{z}(s)$,
\begin{equation}
  \langle f(G) \rangle_{\mathcal{P}} = \int_{\mathbb{H}^{+}} d s \mathcal{W}_{z}(s) f \left(\frac{1}{z - s} \right),
  \label{juda}
\end{equation}  
implying that all moments of the resolvent diagonal elements are determined by the distribution $\mathcal{W}_{z}(s)$ of the self-energy.

The exact Eqs. (\ref{pdfresolven})
and (\ref{pdfresolvencav_Ninf}), albeit having a complicated structure, represent a major step in our understanding of the spectral
properties of random graphs, since they can be solved numerically using a Monte-Carlo iterative method called population
dynamics \cite{Kuhn2008,Rogers2008,Susca2021}. Below we present
an analytic solution of these equations for $c \rightarrow \infty$ and arbitrary degree distributions.

\section{The high-connectivity limit of the resolvent equations}
\label{sec2}

In this section we present a straightforward approach, based on the law of large numbers, to solve the resolvent
equations and determine the spectral and localization properties of random graphs in the high-connectivity limit $c \rightarrow \infty$.
In the appendix, we discuss a more rigorous derivation based on the characteristic functions
of the probability densities $\mathcal{P}_{z}(g)$ and $\mathcal{Q}_{z}(g)$.

Our starting point are the cavity equations (\ref{resolven_eq1}) and (\ref{resolven_eq_cav1}) for a single graph instance, expressed
in terms of the self-energy
\begin{equation}
S_{i}(z) =  \sum_{j \in \partial_i} J_{i j }^2 G_{j j}^{(i)}(z)
\end{equation}
as follows
\begin{equation}
  G_{ii}(z) = \frac{1}{z - S_{i}(z)}, \qquad G_{jj}^{(i)}(z) = \frac{1}{z - S_{j}(z) + J_{ij}^2 G_{ii}^{(j)}(z)  }.
  \label{hsdgal}
\end{equation}  
In the high-connectivity limit $c \rightarrow \infty$, $S_{i}(z)$ is a sum of a large and random number $k_i$ of independent and identically distributed
random variables. By the law of large numbers, $S_{i}(z)$ is asymptotically given by
\begin{equation}
  S_{i}(z) \stackrel{c \rightarrow \infty}{ \longrightarrow} \kappa_i J_1^{2} \langle G \rangle ,
  \label{haso}
\end{equation} 
where $\kappa_i = k_i/c$ is the rescaled degree of node $i$ and the expectation value of $G_{ii}^{(j)}(z)$ is defined as
\begin{equation}
 \langle G \rangle = \int_{\mathbb H^+} dg   \mathcal{Q}_{z}(g) g.
\end{equation}
Therefore, the spatial fluctuations of $S_{i}(z)$ are solely governed by $\kappa_i$ in the limit $c \rightarrow \infty$.
By assuming that the empirical distribution of  $\kappa_1,\dots,\kappa_N$ converges to $\nu(\kappa)$ as $c \rightarrow \infty$,
\begin{equation}
\label{nu_kappa_def}
\nu(\kappa) = \lim_{c\to\infty} \sum_{k=0}^\infty p_k\delta\Big(\kappa - \frac{k}{c}\Big),
\end{equation}
the probability density $\mathcal{W}_z(s)$ of $S_{i}(z)$ is obtained by the change of variables set by Eq. (\ref{haso}), namely
\begin{equation}
  \mathcal{W}_z(s) = \frac{1}{J_{1}^2 {\rm Im} \langle  G \rangle  } \nu \left( \frac{{\rm Im} s}{ J_{1}^2 {\rm Im}  \langle  G \rangle    } \right)
  \delta \left[{\rm Re} s - \frac{{\rm Re} \langle  G \rangle  }{ {\rm Im} \langle  G \rangle  }  {\rm Im} s   \right].
  \label{Ws}
\end{equation}  
We note that $\mathcal{W}_z(s)$
depends itself on the first moment $\langle  G \rangle$ of $G_{ii}^{(j)}(z)$.
This is computed by substituting the  large $c$ behaviour of  $G_{ii}^{(j)}(z)$,
\begin{equation}
G_{ii}^{(j)}(z) \stackrel{c \rightarrow \infty}{ \longrightarrow}  \frac{1}{z -  \kappa_i J_1^{2} \langle G \rangle  },
\end{equation}  
in the definition of the probability density  $\mathcal{Q}_z(g)$ of $G_{ii}^{(j)}(z)$, Eq. (\ref{judadd}), leading to the  self-consistent equation 
\begin{equation}
\mathcal{Q}_z(g) =  \int_{0}^{\infty} d \kappa \, \nu (\kappa) \, \kappa \, \delta \left( g - \frac{1}{z - \kappa J_1^2 \langle G \rangle}    \right).
\end{equation} 
The fixed-point equation for the first moment $\langle G \rangle$ readily follows from the above expression
\begin{equation}
  \langle G \rangle = \int_{0}^{\infty} d \kappa   \frac{  \nu (\kappa) \kappa    }{z - \kappa J_1^2 \langle G \rangle}.
  \label{ffop}
\end{equation} 
Equation (\ref{Ws}) is one of our main analytic results, since the distribution $\mathcal{W}_z(s)$ of the self-energies determines
all moments of the diagonal elements of the resolvent through Eq. (\ref{juda}). Despite the fact we have considered
the high-connectivity limit $c \rightarrow \infty$, the distribution $\mathcal{W}_z(s)$ retains information about
the degree fluctuations through $\nu(\kappa)$. When the tail of the degree distribution $p_k$ decays
fast enough, such as in the case of regular and Erd\"os-R\'enyi random graphs \cite{Newman1}, the rescaled degree distribution
is given by $\nu(\kappa) = \delta(\kappa-1)$, and $\mathcal{W}_z(s)$ reduces to 
\begin{equation}
\mathcal{W}_z(s) = \delta \left( {\rm Im} s -  J_{1}^2 {\rm Im} \langle  G \rangle   \right) \delta \left( {\rm Re} s -  J_{1}^2 {\rm Re} \langle  G \rangle   \right).
\end{equation}  
The above class of solutions describes homogeneous random graphs \cite{Metz2020}, in which the the self-energy $S_i(z)$ is equal
to its mean value $J_1^2 \langle G \rangle$ and the spectral density is given by the Wigner law
\begin{equation}
  \rho_{\rm w} (\lambda) = \frac{\sqrt{4 J_{1}^2 - \lambda^2 }}{2 \pi J_1^2}  \, \boldsymbol{1}_{(-2J_1 ,2J_1)}(\lambda),
  \label{wigner}
\end{equation}
where $\boldsymbol{1}_{\mathcal{A}}(x)$ denotes the indicator function, i.e., $\boldsymbol{1}_{\mathcal{A}}(x) = 1$ if $x \in \mathcal{A}$, and
 $\boldsymbol{1}_{\mathcal{A}}(x) = 0$ otherwise. 

Let us derive some consequences of Eq. (\ref{Ws}).
By inserting Eq. (\ref{Ws}) in (\ref{juda}) and then making
a change of integration variables, we obtain 
\begin{equation}
\label{rho_nu}
\rho_{\epsilon}(\lambda) = \frac{1}{\pi}  \mathrm{Im}\Bigg[ \int_{0}^{\infty} d\kappa\frac{\nu(\kappa)}{z- \kappa J_1^2 \langle G \rangle }\Bigg]
\end{equation}
and
\begin{equation}
\label{P_nu}
 \mathcal{I}_{\epsilon} (\lambda) =  \frac{\epsilon}{\pi \rho_{\epsilon}(\lambda)} \int_{0}^{\infty} d\kappa\frac{\nu(\kappa)}{|z- \kappa J_1^2 \langle G \rangle|^2}
\end{equation}
from Eqs. (\ref{rho_1}) and (\ref{P_1}).
We can also derive the analytic expression for the joint distribution $\mathcal{P}_z(g)$ of the
diagonal elements of the resolvent. Let $G(z)$ and $S(z)$ be independent complex random variables distributed
according to $\mathcal{P}_z(g)$ and $\mathcal{W}_z(s)$, respectively, then Eq. (\ref{hsdgal}) entails
\begin{equation}
G(z) \stackrel{d}{=} \frac{1}{z-S(z)}.
\end{equation}  
By making a two-dimensional change of variables and using Eq. (\ref{Ws}), we find
\begin{equation}
  \fl
  \mathcal{P}_z(g) = \frac{1}{J_1^2 |g|^4 {\rm Im} \langle G \rangle} \nu \left( \frac{  {\rm Im}  \left( z - g^{-1}  \right)  }{ J_1^2   {\rm Im} \langle G \rangle }  \right)
  \delta \left[ {\rm Re}  \left( z - g^{-1}  \right) - \frac{{\rm Re} \langle G \rangle }{ {\rm Im} \langle G \rangle }  {\rm Im}  \left( z - g^{-1}  \right)   \right].
  \label{hdsk}
\end{equation}  
The above equation determines how the distribution of the diagonal part of the resolvent depends on the distribution $\nu(\kappa)$ of rescaled
degrees. The object $\mathcal{P}_z(g)$ contains much information about the spectral and localization properties
of the adjacency matrix \cite{Tikhonov2019,Biroli2022,Tapias2022}. For instance, by marginalizing $\mathcal{P}_z(g)$ with respect to ${\rm Re} g$, we can calculate the empirical
distribution of $\{ {\rm Im} G_{ii} \}_{i=1,\dots,N}$, which is essentially the (regularized) local density of states \cite{Mirlin2000,Tikhonov2019,Biroli2022}
\begin{equation}
  \rho_i(z) = \frac{1}{\pi} {\rm Im} G_{ii}(z) = \frac{1}{\pi} \sum_{\mu=1}^N \frac{\epsilon |v_{\mu,i} |^2  }{\left( \lambda -\lambda_{\mu}   \right)^2 + \epsilon^2 }   \quad (i=1,\dots,N).
  \label{hhssl}
\end{equation}  
In the limit $\epsilon \rightarrow 0^{+}$, the empirical distribution
of $y_i = {\rm Im} G_{ii}(z)$ ($i=1,\dots,N$) characterizes the spatial fluctuations of the eigenvector
amplitudes $|v_{\mu,i}|^2$ corresponding to the eigenvalues around $\lambda$.
By  integrating Eq. (\ref{hdsk}) over ${\rm Re} g$, we find the expression for $|\lambda|>0$ 
\begin{eqnarray}
  P_z(y) &= \Bigg{\{} \omega_{+}(y) \, \nu\left( \frac{y \left[ x_{+}^2(y) + y^2   \right]^{-1} - \epsilon  }{J_1^{2} \, {\rm Im} \langle G \rangle  }  \right) \nonumber \\
  & + \omega_{-}(y) \, \nu\left( \frac{y \left[ x_{-}^2(y) + y^2   \right]^{-1} - \epsilon  }{J_1^{2} \, {\rm Im} \langle G \rangle  }  \right) \Bigg{\}} \boldsymbol{1}_{(0,y_{\rm e})}(y),
  \label{huda33}
\end{eqnarray}  
where the support of $P_z(y)$ is determined by
\begin{equation}
 y_{\rm e} =  \frac{|{\rm Re} \langle G \rangle  | + \sqrt{\left(  {\rm Re} \langle G \rangle  \right)^2 + \left(  {\rm Im} \langle G \rangle  \right)^2}}
 {2 \left( |\lambda|   {\rm Im} \langle G \rangle  + \epsilon \, |{\rm Re} \langle G \rangle|   \right)  }.
 \label{support}
\end{equation}  
The functions $\omega_{\pm}(y)$ and $x_{\pm}(y)$ are defined as
\begin{equation}
\omega_{\pm}(y) = \frac{1}{J_1^{2} \, {\rm Im} \langle G \rangle  \Big{|} x_{\pm}^2(y) - y^2 + 2 \frac{{\rm Re} \langle G \rangle }{ {\rm Im} \langle G \rangle } \, y \,  x_{\pm}(y)   \Big{|} }
\end{equation}
and
\begin{equation}
  x_{\pm}(y) = \frac{1 \pm  \sqrt{1 - 4 \left(  \lambda + \frac{{\rm Re} \langle G \rangle }{ {\rm Im} \langle G \rangle } \epsilon  \right)
      \left[ \left(  \lambda + \frac{{\rm Re} \langle G \rangle }{ {\rm Im} \langle G \rangle } \epsilon  \right) y^2 - \frac{{\rm Re} \langle G \rangle }{ {\rm Im} \langle G \rangle } y   \right]}}
  { 2 \left(   \lambda + \frac{{\rm Re} \langle G \rangle }{ {\rm Im} \langle G \rangle } \epsilon  \right)   }.
\end{equation}  
Equation (\ref{huda33}) shows that $P_z(y)$ has a finite support for $|\lambda| >0$ regardless of the shape of the distribution $\nu$.
By setting $\lambda=0$ in Eq. (\ref{hdsk}) and then integrating over ${\rm Re} g$, we obtain
\begin{equation}
  P_z(y) = \frac{1}{J_1^2 {\rm Im} \langle G \rangle y^2 } \nu \left( \frac{- \epsilon + y^{-1}  }{J_1^2 {\rm Im} \langle G \rangle   }   \right)
  \quad (\lambda=0).
  \label{huda34}
\end{equation}  
In contrast to Eq. (\ref{huda33}), the distribution $P_z(y)$  at $\lambda=0$ can have an infinite support depending
on the choice of the rescaled degree distribution $\nu$.

\section{Results for the negative binomial degree distribution}
\label{sec3}

The analytic expressions of the previous section are valid for any distribution
of rescaled degrees $k_i/c$ that converges to $\nu(\kappa)$ as $c \rightarrow \infty$.
Here we discuss results for random graphs with the
negative binomial degree distribution \cite{evans2011statistical}
\begin{equation}
\label{neg_bin1}
     p_k^{\mathrm{(b)}} = \frac{\Gamma (\alpha + k)}{\Gamma(\alpha)}\frac{1}{k!}\left(\frac{c}{\alpha}\right)^k \frac{1}{(1 + \frac{c}{\alpha})^{\alpha +k}},
\end{equation}
where $\Gamma(x)$ is the Gamma function and $0 < \alpha < \infty$ is a continuous parameter.
The variance $\sigma_{b}^{2}$ of $p_k^{\mathrm{(b)}}$ is related to $\alpha$ as follows 
\begin{eqnarray}
 \label{polaq}
\sigma_{b}^{2} = c + \frac{c^2}{\alpha}.
\end{eqnarray}
In a previous work \cite{Metz2020}, we have shown that spectral density of the configuration model does not converge to
the Wigner law if the relative variance of the degree distribution does not vanish as $c \rightarrow \infty$.
The negative binomial degree distribution provides a controllable way to investigate the effect of degree fluctuations on the spectral
and localization properties of random graphs by varying a single parameter. In fact, given that
\begin{equation}
  \lim_{c \rightarrow \infty} \frac{\sigma_{b}^{2}}{c^2} = \frac{1}{\alpha},
  \label{gdja}
\end{equation}  
by changing $\alpha$ we are able to explore the entire range of degree fluctuations for $c \rightarrow \infty$. The limit
$\alpha \rightarrow \infty$ corresponds to homogeneous random graphs, whose spectral properties are governed by random
matrix theory \cite{Livan2018}, whereas the limit $\alpha \rightarrow 0$ characterizes random graphs with strongly heterogeneous degrees.
The geometric degree distribution is recovered for $\alpha = 1$ \cite{Metz2020}.
Inserting Eq. (\ref{neg_bin1}) in Eq. (\ref{nu_kappa_def}), we obtain the analytic form of $\nu(\kappa)$
\begin{eqnarray}
\label{nu_neg_binomial}
    \nu_{b} (\kappa) = \frac{\alpha^\alpha \kappa^{\alpha - 1} e^{-\alpha\kappa}}{\Gamma(\alpha)}.
\end{eqnarray}
The above expression is the only input to the general formulae of the previous section, from which we
can derive several analytic results as a function of $\alpha$.


\subsection{Spectral density}

In the high-connectivity limit, the empirical
distribution  $\nu(\kappa)$ of rescaled degrees determines the spectral density $\rho(\lambda)$, as the latter
is given by the free multiplicative convolution of $\nu(\kappa)$ with the Wigner law $\rho_{\rm w}(\lambda)$. This rigorous result, proven
in \cite{dembo2021empirical}, essentially means that the adjacency matrix can be decomposed in the limit $c \rightarrow \infty$ as a product of
$\boldsymbol{X}$ and $\boldsymbol{D}$, where
$\boldsymbol{D}$ is the degree matrix with elements $D_{ij} = \kappa_i \delta_{ij}$, and $\boldsymbol{X}$ is a random matrix in which
the diagonal entries are zero and the off-diagonal elements  are independent random variables drawn from a Gaussian distribution with mean $J_0/N$ and variance $J_{1}^2/N$.
The product $\boldsymbol{X}\boldsymbol{D}$, however, is non-Hermitian and its eigenvalues could be complex numbers. Fortunately, $\boldsymbol{D}$ is a positive operator and
$\boldsymbol{D}^{1/2} \boldsymbol{X} \boldsymbol{D}^{1/2}$ is Hermitian, with the same moments as $\boldsymbol{X}\boldsymbol{D}$, which
allows us to rewrite the adjacency matrix as
\begin{eqnarray}
\label{decomp}
\boldsymbol{A} = \boldsymbol{D}^{1/2} \boldsymbol{X} \boldsymbol{D}^{1/2}.
\end{eqnarray}
This interesting decomposition can be used to study the spectral properties of random
graphs with  a prescribed degree distribution and large $c$ without having to run sophisticated
algorithms to sample graphs from the configuration model.
This is precisely the strategy we adopt below, i.e., we compare our theoretical findings with
numerical results obtained from diagonalizing Eq. (\ref{decomp}). 
\begin{figure}
    \centering
    \includegraphics[width=0.43\linewidth]{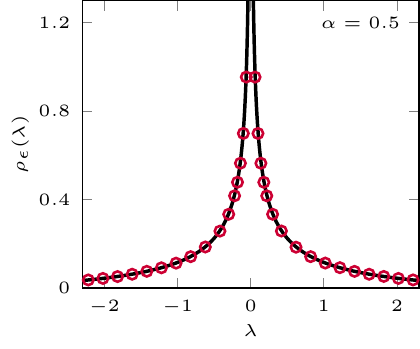}
    \includegraphics[width=0.43\linewidth]{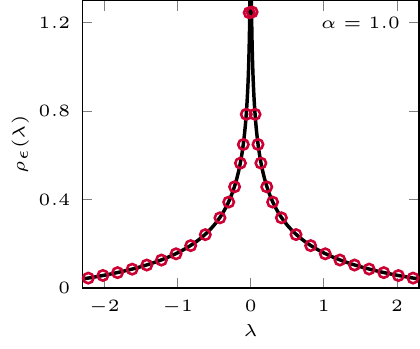} \\
     \includegraphics[width=0.43\linewidth]{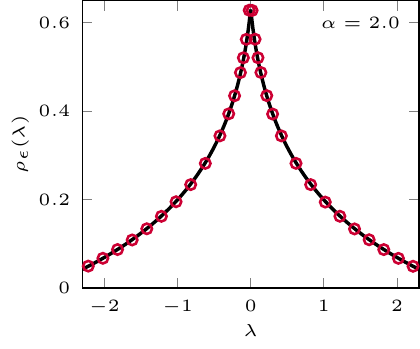}
    \includegraphics[width=0.43\linewidth]{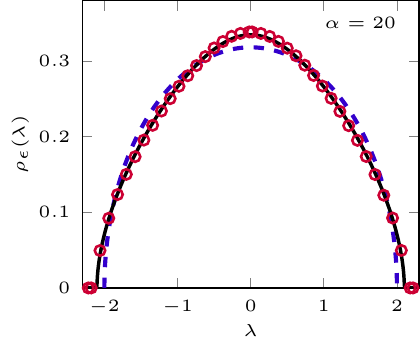}
  \caption{The spectral density of random graphs with a negative binomial degree distribution in the high-connectivity limit. The
    parameter $1/\alpha$ controls the relative variance of the degree distribution (see Eq. (\ref{gdja})). The solid lines are the theoretical
    results derived from solving Eqs. (\ref{Gcav_neg_binomial}) and (\ref{rho_neg_binomial}) for $\epsilon=10^{-3}$ and $J_1=1$. The red circles are numerical diagonalization
    results obtained from an ensemble of $10^4\times 10^4$ adjacency random matrices generated according to Eq. (\ref{decomp}). The dashed blue curve
    in the lower right panel represents the Wigner law (see Eq. (\ref{wigner})).
  }
  \label{Fig1}
\end{figure}

Let us determine the spectral density of random graphs with a negative binomial degree distribution. Substituting Eq. (\ref{nu_neg_binomial})
in Eqs. (\ref{ffop}) and (\ref{rho_nu}), and evaluating the integrals over $\kappa$, we obtain
\begin{equation}
\label{rho_neg_binomial}    
     \rho_{\epsilon}(\lambda) = \frac{1}{\pi} \mathrm{Im}\Bigg[\frac{z^2 + \gamma^2J_1^2}{z\gamma^2J_1^2}\Bigg],
\end{equation}
where the dimensionless variable $\gamma \in \mathbb{C}$, defined in terms of $\langle G \rangle$ as
\begin{equation}
\gamma = \frac{z}{J_1^2 \langle G \rangle},
\end{equation} 
solves the transcendental equation
\begin{equation} 
    \label{Gcav_neg_binomial}
     \gamma^2 J_1^2= \frac{z^2}{ \left(-\alpha \gamma e^{-\gamma}   \right)^{\alpha}   \Gamma (1-\alpha,-\alpha \gamma) - 1},
\end{equation}
with $\Gamma(a,\xi)$ ($a \in \mathbb R$ and $\xi \in \mathbb C$) denoting the incomplete Gamma function.
The solution of the fixed-point Eq. (\ref{Gcav_neg_binomial}) yields the regularized spectral density (\ref{rho_neg_binomial}) for
any $0 < \alpha < \infty$. We recall that the strength of the degree fluctuations
is controlled only by $\alpha$ (see Eqs. (\ref{polaq}) and (\ref{gdja})).  
By setting $\alpha=1$ in Eq. (\ref{Gcav_neg_binomial}), we
recover the equations for the spectral
density of random graphs with a geometric degree distribution \cite{Metz2020}.

Figure \ref{Fig1} compares the regularized spectral density $\rho_{\epsilon}(\lambda)$ computed from the solutions of Eq. (\ref{Gcav_neg_binomial}) with
numerical results for the eigenvalues obtained from diagonalizing the adjacency matrix of Eq. (\ref{decomp}).
The agreement between our theoretical findings and numerical diagonalization results is excellent. In particular, we note from figure \ref{Fig1} that degree fluctuations
modify the tails of the spectral density as well as its behaviour around $\lambda=0$.

We have shown in a previous work \cite{Metz2020} that $\rho(\lambda)$ has a logarithmic divergence at $\lambda=0$ for $\alpha=1$. In order
to understand how this singular behaviour depends on $\alpha$, we need to extract the functional form of $\gamma=\gamma(z)$ as $|z| \to 0$.
We follow \cite{Metz2020} and make the assumption
\begin{equation}
\label{gamma_ansatz}
\gamma(z) = \frac{\beta_1}{J_1} z + \frac{\beta_2(\alpha,z)}{ J_{1}^2 }z^2,
\end{equation}
where the coefficient $\beta_1$ is independent of $z$ and $\beta_2(\alpha,z)$ satisfies $\lim_{|z|\to 0} z^2\,\beta_2(\alpha,z) = 0$.
Inserting the above {\it ansatz} in Eq. (\ref{Gcav_neg_binomial}) and expanding the result up to $\mathcal{O}(z^2)$, one finds
that $\beta_1$ and $\beta_2(\alpha,z)$ are given by
\begin{equation}
\label{beta1}
\beta_1 = - i
\end{equation}
and 
\begin{equation}
\label{beta2}
\beta_2(\alpha,z) = -\frac{1}{2}\Bigg[i^{2\alpha}\alpha^\alpha \bigg(\frac{\beta_1}{J_1}z\bigg)^{\alpha-1}\Gamma(1-\alpha) + \frac{\alpha}{1-\alpha}\Bigg] 
\end{equation}
in the regime $\alpha \in (0,1)$.

\begin{figure}
    \centering
  \includegraphics[width=0.62\linewidth]{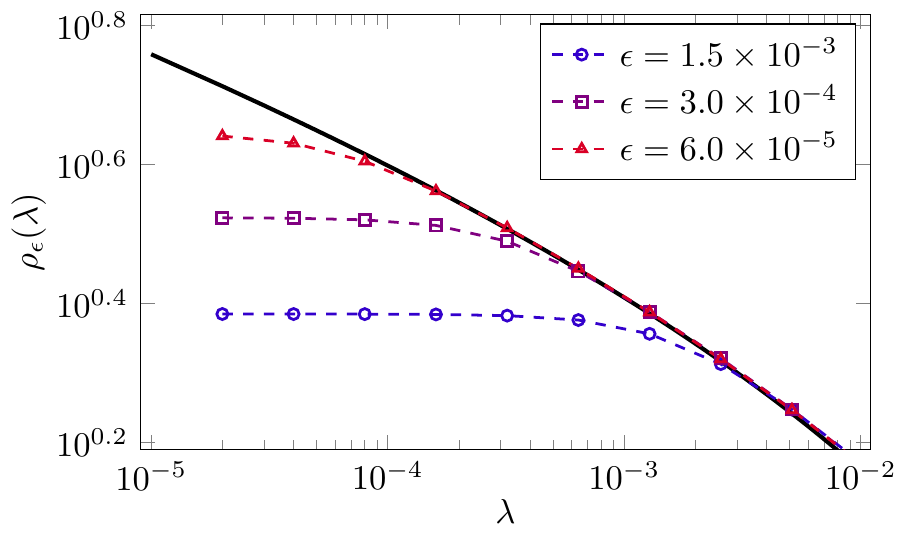}
  \caption{The power-law divergence of the spectral density around $\lambda = 0$ for $\alpha=0.9$ and $J_1=1$. The solid
    line is the analytic result of Eq. (\ref{rho_small_lambda}), while the symbols are numerical results obtained
    from the solutions of Eqs. (\ref{Gcav_neg_binomial}) and (\ref{rho_neg_binomial}) for different values of $\epsilon$.
    The data is presented in logarithmic scale.
  }
  \label{Fig2}
\end{figure}

The last step is to substitute Eq. (\ref{gamma_ansatz}) in Eq. (\ref{rho_neg_binomial}) and compute the limit $\epsilon \to 0^+,$ which leads to the power-law divergence
\begin{equation}
  \label{rho_small_lambda}
  \fl
\rho(\lambda) = \frac{1}{\pi J_1} \Bigg[ \alpha^{\alpha} \sin\bigg(\frac{\pi}{2}\alpha\bigg)\Gamma(1-\alpha) \left( \frac{|\lambda|}{J_1} \right)^{\alpha-1} - \frac{\alpha}{(1-\alpha)}\Bigg]
\quad (0 < \alpha < 1)
\end{equation}
for $|\lambda|\to 0$.
By taking the limit $\alpha \rightarrow 1$ in Eq. (\ref{rho_small_lambda}), we recover the logarithmic divergence obtained in \cite{Metz2020}
\begin{equation}
\fl
\rho(\lambda) = -\frac{1}{\pi J_1}\Bigg[E + \log \left( \frac{|\lambda|}{J_1} \right) \Bigg]
\quad (\alpha=1),
\end{equation}
with $E$ representing the Euler-Mascheroni constant. Figure \ref{Fig2} compares Eq. (\ref{rho_small_lambda}) with
numerical solutions of Eqs. (\ref{Gcav_neg_binomial}) and (\ref{rho_neg_binomial}) for $|\lambda| \ll 1$. The numerical results deviate from the
analytic expression for values of $\lambda$ below a certain threshold $|\lambda_*|$. As $\epsilon$ decreases, $|\lambda_*|$ shifts towards smaller values, confirming
that the discrepancy between the numerical data and Eq. (\ref{rho_small_lambda}) is due to the finite values of $\epsilon$ used in the numerical
solutions.

In the homogeneous limit $\alpha \rightarrow \infty$, the variance of the rescaled degree distribution $\nu(\kappa)$ vanishes and we expect
to recover the Wigner law.
By using the functional relation \cite{grad2014}
\begin{equation}
\Gamma (1-\alpha,-\alpha \gamma) = - \alpha \Gamma (-\alpha,-\alpha \gamma) + \left(- \alpha \gamma   \right)^{-\alpha} e^{\alpha \gamma}
\end{equation}  
and the asymptotic formula \cite{Nemes2016}
\begin{equation}
  \fl
  - \alpha \left( - \alpha \gamma  \right)^{\alpha} e^{-\alpha \gamma}  \Gamma (-\alpha,-\alpha \gamma) =
  \frac{1}{ \left( \gamma -1 \right)} + \frac{\gamma}{ \alpha \left( \gamma -1 \right)^3} + \mathcal{O}(\alpha^{-2}) \quad  (\alpha \gg 1),
\end{equation}  
we derive from Eq. (\ref{Gcav_neg_binomial}) an approximate equation for $\gamma$ 
\begin{equation}
  \gamma^2 J_1^2  = z^2 (\gamma -1) - \frac{z^2 \gamma }{\alpha (\gamma -1)  }.
  \label{hfkskd}
\end{equation}  
In the limit $\alpha  \rightarrow \infty$, the above expression reduces to a quadratic equation, whose solution yields the Wigner law (see Eq. (\ref{wigner})).

\begin{figure}
    \centering
  \includegraphics[width=0.59\linewidth]{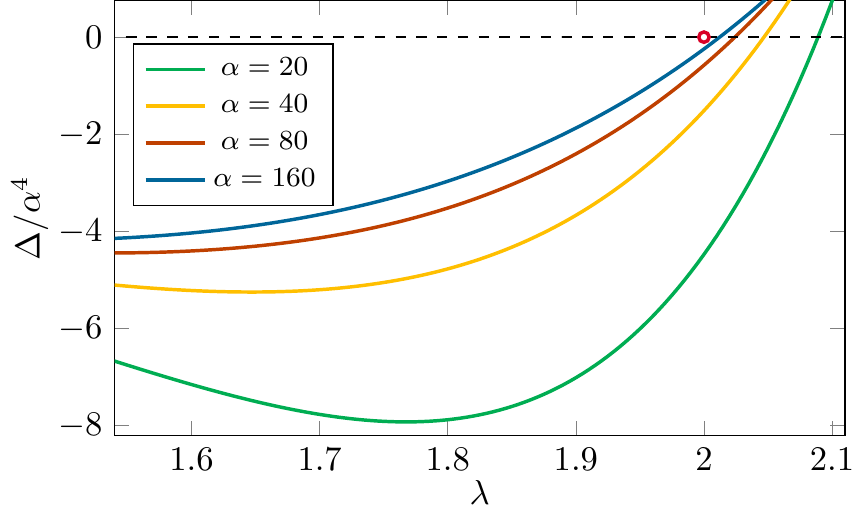}
  \caption{The discriminant $\Delta(\lambda)$ of the cubic Eq. (\ref{hfkskd}) for $J_1=1$ and $\epsilon=0$. The edge of the spectral density is determined
    by the value of $\lambda_{\rm b}$ at which $\Delta(\lambda_{\rm b} ) =0$. The red circle identifies the spectral edge $\lambda_{\rm b}=2 J_1$ of the Wigner law (see Eq. (\ref{wigner})).}
  \label{Fig2das}
\end{figure}
Here we do not derive the analytic expression for the spectral density $\rho(\lambda)$ that arises from solving the cubic Eq. (\ref{hfkskd}), but
we characterize the support of $\rho(\lambda)$, which plays a pivotal role for the stability
of complex systems \cite{Neri2020}.
In general, when the largest eigenvalue of the adjacency matrix $\boldsymbol{A}$ is finite, there exists a regime of model
parameters where the stationary states of a large complex system coupled through $\boldsymbol{A}$ are linearly stable.
Thus, complex systems interacting through the symmetric random matrix of Eq. (\ref{decomp}) can be in a stable state in
the limit $\alpha \rightarrow \infty$, in view of the finite support of the Wigner
law. An interesting question here is whether the support of $\rho(\lambda)$ remains finite when
a small amount of heterogeneity is introduced ($1 \ll \alpha < \infty$).
In order to resolve this issue, we study the discriminant $\Delta(\lambda)$ of
the cubic Eq. (\ref{hfkskd}) in the limit $\epsilon \rightarrow 0^{+}$.
If $\Delta(\lambda) > 0$, then  Eq. (\ref{hfkskd}) has only real roots and $\rho(\lambda)=0$, whereas if  $\Delta(\lambda) < 0$, then
Eq. (\ref{hfkskd}) admits a pair of complex-conjugate solutions, yielding $\rho(\lambda)>0$. 
As shown in figure \ref{Fig2das}, the discriminant is zero at a certain value $\lambda=\lambda_{\rm b}$, which implies
that $\rho(\lambda)$  has a finite support. The spectral edge $\lambda_{\rm b}$  of $\rho(\lambda)$
consistently approaches the value $\lambda_{\rm b} = 2 J_1 $ of the Wigner law as $\alpha$ increases.


\subsection{Eigenvector localization and the distribution of the local density of states}

In this section we analyse the effect of degree fluctuations on the inverse participation ratio (IPR) and on the
local density of states (LDOS) for a negative binomial degree distribution.
Substituting Eq. (\ref{nu_neg_binomial}) in Eq. (\ref{P_nu}) and calculating the integral over $\kappa$, we obtain
the regularized IPR around an eigenvalue $\lambda$ 
\begin{eqnarray}
  \label{IPRneg}
  \fl
  & \mathcal{I}_{\epsilon} (\lambda) = \frac{\epsilon}{\pi \rho_\epsilon(\lambda)}\mathrm{Im}\Bigg\{  \frac{\alpha^\alpha \gamma}
     {z \left[ \mathrm{Im} \left( z/\gamma \right)  \right]^{\alpha-1}
    \left[ \epsilon+ \gamma \, \mathrm{Im} \left( z/\gamma  \right)  \right] } \times  \\
    \fl
&  \Bigg[ \epsilon^{\alpha-1}\exp\left( \frac{\epsilon \, \alpha}{\mathrm{Im} \left( z/\gamma  \right)}\right)
      \Gamma\left(1-\alpha,\frac{\epsilon \, \alpha}{\mathrm{Im} \left(z/\gamma\right)}\right)    - \left[ -\gamma \mathrm{Im} \left( z/\gamma \right)   \right]^{\alpha-1}
      e^{-\alpha\gamma}  \Gamma \left( 1-\alpha,-\alpha\gamma \right) \Bigg]\Bigg\}, \nonumber 
\end{eqnarray}
where $\gamma$ fulfills Eq. (\ref{Gcav_neg_binomial}) and the regularized spectral density $\rho_\epsilon(\lambda)$ is
given by Eq. (\ref{rho_neg_binomial}).

\begin{figure}[H]
    \centering
  \includegraphics[width=1.0\linewidth]{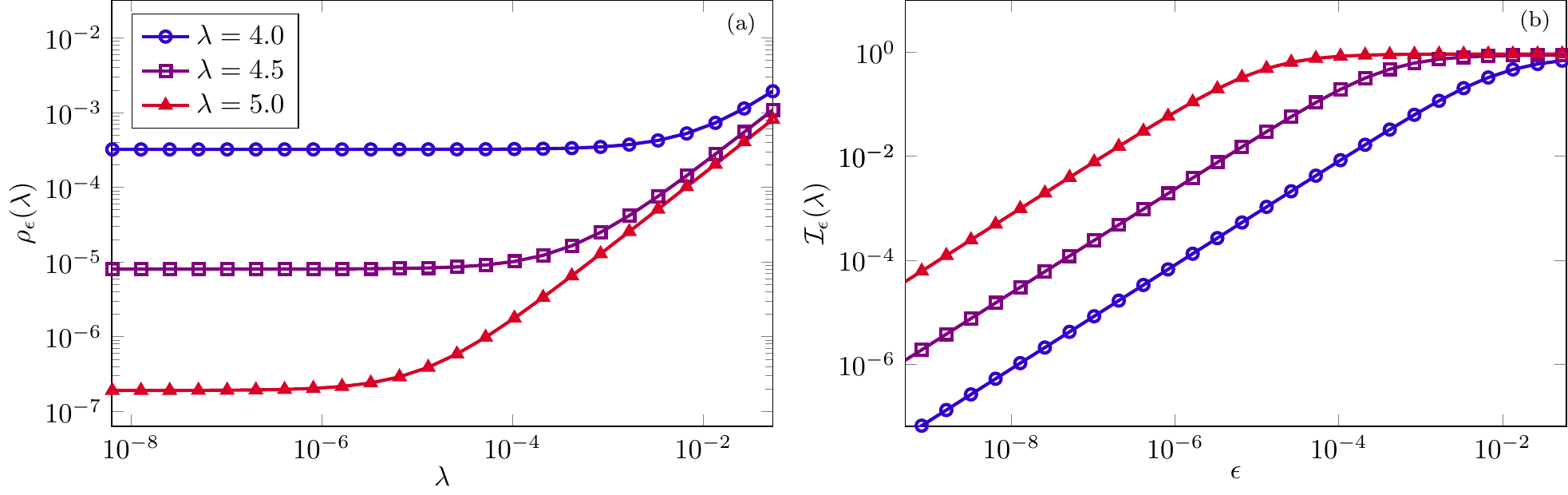}
  \caption{The regularized spectral density $\rho_{\epsilon}(\lambda)$  and the inverse participation ratio $\mathcal{I}_{\epsilon}(\lambda)$ as a function of the
    regularization parameter $\epsilon$ for $\alpha=0.75$ and different values of $\lambda$. The numerical results are
    obtained from Eqs (\ref{rho_neg_binomial}) and (\ref{IPRneg}).}
  \label{Fig3}
\end{figure}
It is well-established that the eigenvectors of random graphs with finite $c$ become localized in the tails of the spectral
density due to the existence of hubs in the graph structure \cite{Metz2010,Goltsev2012,Slanina2012}. It is natural to ask whether such localized
states survive for $c \rightarrow \infty$ in the presence of degree fluctuations. Figure \ref{Fig3} shows the spectral density and
the IPR derived from Eqs. (\ref{rho_neg_binomial}) and (\ref{IPRneg}) as a function of $\epsilon$ for large values of $|\lambda|$. 
In the regime $\epsilon \rightarrow 0^+$, the spectral density $\rho_{\epsilon}(\lambda)$ converges to a finite limit and the IPR
vanishes as $\mathcal{I}_{\epsilon}(\lambda) \propto \epsilon$.
The same picture holds for other values of $\alpha$, which demonstrates that all
eigenvectors corresponding to nonzero eigenvalues are extended.

\begin{figure}[H]
    \centering
     \includegraphics[width=0.98 \linewidth]{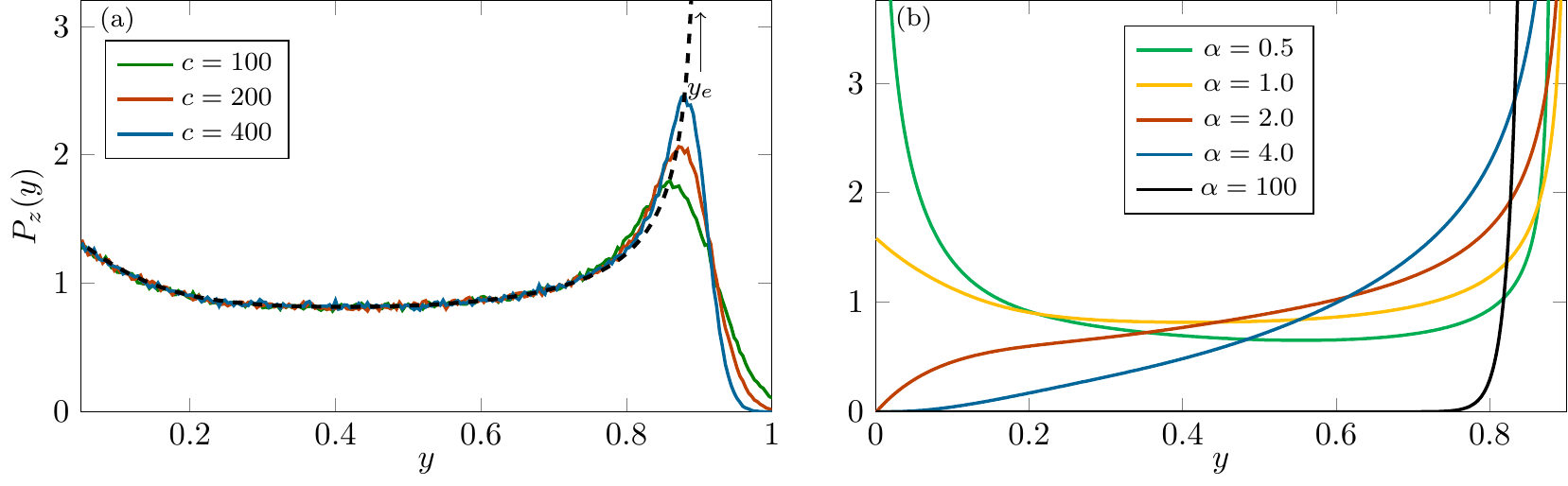}
  \caption{The empirical distribution of the imaginary part of the resolvent at $(\lambda,\epsilon) = (1,10^{-3})$ for random graphs with $J_1=1$ and a negative
    binomial degree distribution (see Eq. (\ref{nu_neg_binomial})). (a) Results for $\alpha=1$. The dashed line
    is the analytic expression of Eq. (\ref{huda33}), valid in the high-connectivity limit $c \rightarrow \infty$, while the solid lines are obtained from the numerical solutions
    of Eqs. (\ref{pdfresolven}) and (\ref{pdfresolvencav_Ninf}) for different values of the mean degree $c$. The support of the distribution
    is limited by $y_{\rm e}$ (see Eq. (\ref{support})). (b) The analytic result
    of Eq. (\ref{huda33}) for different values of $\alpha$.    
  }
  \label{FigLDOS}
\end{figure}

The distribution $P_z(y)$ of the LDOS probes the spatial fluctuations of the eigenvectors and it
gives important information about localization phenomena.
In the limit $\epsilon \rightarrow 0^{+}$, the distribution $P_z(y)$ within the localized phase typically
exhibits a singularity at $y \simeq \epsilon$, due to the extensive number of 
sites at which ${\rm Im} G_{ii} \simeq \epsilon$ \cite{Abou1973}. Differently from that, our results show that $P_z(y)$ converges to a regular, $\epsilon$-independent function
in the limit $\epsilon \rightarrow 0^{+}$, highlighting the extended nature of the eigenvectors. Figure \ref{FigLDOS}-(a) compares Eq. (\ref{huda33}) with numerical
results obtained from the solutions of Eqs. (\ref{pdfresolven}) and (\ref{pdfresolvencav_Ninf}) using the population dynamics algorithm \cite{Reimer2011}
for $\alpha=1$ and large values of $c$. The agreement between theoretical and numerical results is excellent over the central
portion of the distribution. The discrepancy close to $y = y_{\rm e}$ in figure \ref{FigLDOS}-(a) is due to strong
finite-connectivity effects, since the convergence of the  numerical results to the asymptotic behaviour for $c \rightarrow \infty$ is extremely slow.

Figure \ref{FigLDOS}-(b) shows that $P_z(y)$ diverges at the edge $y=y_{\rm e}$ for any
value of $\alpha$. Moreover, the distribution $P_z(y)$ develops an additional power-law singularity at $y=0$ when $\alpha < 1$, which is a genuine effect of strong degree
fluctuations and a direct consequence of the shape of $\nu$ (see Eq. (\ref{nu_neg_binomial})).
In the limit $\alpha \rightarrow \infty$, the
graph becomes homogeneous and $P_z(y)$ converges to a Dirac-$\delta$ distribution
centered at $y_{\rm e} = \pi \rho_{\rm w}(\lambda)$ ($\epsilon \rightarrow 0^{+}$), where $\rho_{\rm w}(\lambda)$ is given by Eq. (\ref{wigner}).

\begin{figure}[H]
    \centering
    \includegraphics[width=0.99 \linewidth]{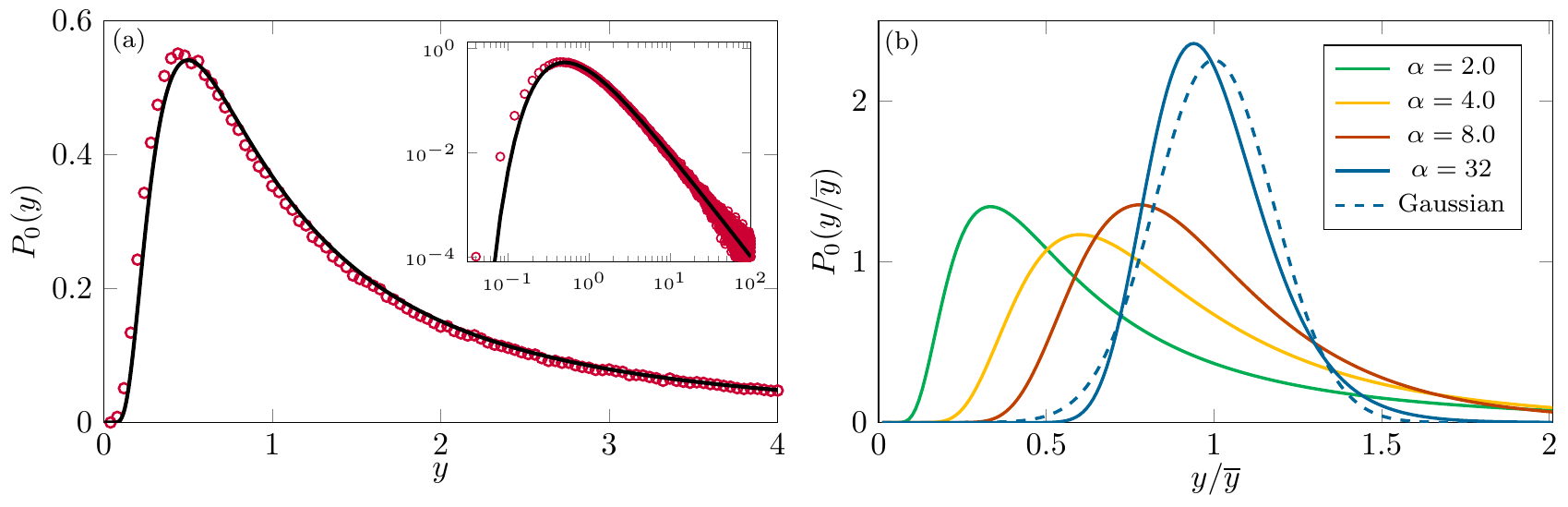}
  \caption{The empirical distribution of the imaginary part of the resolvent at $z=0$ for random graphs with $J_1=1$ and a negative
    binomial degree distribution (see Eq. (\ref{nu_neg_binomial})). (a) Results for $\alpha=1$. The solid line
    is the analytic result of Eq. (\ref{hkskd}), while the red circles are obtained from the numerical solutions
    of Eqs. (\ref{pdfresolven}) and (\ref{pdfresolvencav_Ninf}) for $(\lambda,\epsilon) = (0,10^{-3})$ and $c=400$. The inset shows
    the tail of the distribution in logarithmic scale. (b) The distribution of $y$  rescaled by its mean value $\overline{y}$ for different $\alpha$ (see Eqs. (\ref{hkskd})
    and (\ref{momo})). The dashed line is a Gaussian distribution with unity mean and variance $1/\alpha$ ($\alpha=32$).   
  }
  \label{FigLDOS1}
\end{figure}

With the aim of clarifying the singular behaviour of the spectral density (see figure \ref{Fig2}), we turn our attention to the statistics of the LDOS at $\lambda=0$.
Substituting Eq. (\ref{nu_neg_binomial}) in Eq. (\ref{huda34}) and setting  $\epsilon = 0$, we obtain the simple analytic result
\begin{equation}
  P_{0}(y) = \frac{\alpha^{\alpha}}{\Gamma(\alpha) J_{1}^{\alpha}} \frac{e^{- \frac{\alpha}{J_1 y}   }}{y^{\alpha+1}},
  \label{hkskd}
\end{equation}
which reveals the unbounded character of the LDOS fluctuations at $\lambda=0$.
Figure (\ref{FigLDOS1})-(a) confirms the exactness of  expression (\ref{hkskd}) by comparing this equation with results
obtained by numerically solving Eqs. (\ref{pdfresolven}) and (\ref{pdfresolvencav_Ninf})
for $\alpha=1$ and $c=400$. It is interesting to contrast $P_{0}(y)$
with the distribution of the LDOS in the extended
phase of regular random graphs with on-site random potentials \cite{Mirlin1994,Mirlin1994A,Biroli2022}.
While for regular random graphs with on-site disorder the distribution of the LDOS decays exponentially
fast beyond a certain scale \cite{Mirlin1994,Mirlin1994A}, the power-law tail of Eq. (\ref{hkskd})
implies that the $q$-th moment $\overline{y^q} = \int_{0}^{\infty} d y \, y^q \, P_{0}(y)$ diverges
for $\alpha \leq q$, whereas
\begin{equation}
  \overline{y^q} = \frac{\alpha^q \Gamma \left( \alpha - q \right)}{J_{1}^q \Gamma(\alpha) }
  \label{momo}
\end{equation}
for $\alpha > q$.
Figure (\ref{FigLDOS1})-(b) shows that $\overline{y}$ does not coincide with the most probable value of
the distribution $P_{0}(y)$ due to its skewed shape.
As $\alpha$ increases and the graph
becomes more homogeneous, the distribution $P_{0}(y/\overline{y})$ gradually becomes more symmetric and concentrated around its mean value.
For $1 \ll \alpha < \infty$,  $P_{0}(y/\overline{y})$ is a Gaussian distribution
with variance $\mathcal{O}(1/\alpha)$, and it ultimately converges to $P_{0}(y/\overline{y}) = \delta(y/\overline{y} -1)$ in the homogeneous limit $\alpha \rightarrow \infty$.


\section{Summary and discussion}
\label{sec4}

The resolvent distributional equations for the spectral properties of heterogeneous random graphs do not have
analytic solutions for finite mean degree $c$. In the limit $c \rightarrow \infty$, such
equations admit a trivial solution, typical of random graphs with a homogeneous structure, in which the resolvent elements
are all equal to their mean value.
Here we have shown how to distill a nontrivial analytic solution of the resolvent
distributional equations, valid in the high-connectivity
limit, which explicitly depends on the shape of the degree distribution. This solution enables
to perform a thorough analysis of the impact of degree heterogeneities on the spectral and localization properties of
the adjacency matrix. 

We have presented several results for the spectral and localization properties of
random graphs with a negative binomial degree distribution, in which
the network heterogeneity, measured by the relative variance of the degree distribution (see Eq. (\ref{gdja})), is governed by a single parameter $\alpha \in (0,\infty)$.
When the degree fluctuations are sufficiently strong ($0 < \alpha \leq 1$), the spectral density $\rho(\lambda)$
diverges at the zero eigenvalue $\lambda=0$. More specifically, the function $\rho(\lambda)$ exhibits either a logarithmic or a power-law
singularity if $\alpha=1$ or $\alpha \in (0,1)$, respectively.
In addition, we have shown that $\rho(\lambda)$ has a finite support in the regime of weak degree fluctuations ($1 \ll \alpha < \infty$), which implies
that large complex systems coupled
through  highly connected random graphs can be found in a linearly stable state \cite{Neri2020}, at least when the variance
of the degree distribution is small enough.
An interesting open question is whether $\rho(\lambda)$ becomes unbounded below
a critical value of $\alpha$, or if the support of $\rho(\lambda)$ remains always finite.

We have shown that the inverse participation ratio vanishes for nonzero eigenvalues and the corresponding eigenvectors are extended for any
amount of degree fluctuations.
We point out that this picture is not in conflict with recent results \cite{Alt2021,Tarzia2022} that
show the existence of localized eigenvectors in the tails of the spectral density of critical random
graph models. In fact, our results for the absence of localization hold for $c=\mathcal{O}(N^{a})$ ($a < 1$) \cite{Metz2022}, while
in critical random graphs the mean degree scales as $c=\mathcal{O}(\ln N)$. 

In order to further examine the nature of the eigenvectors and the singular behaviour of the spectral density, we have computed analytically the distribution
of the local density of states (LDOS), which quantifies the spatial fluctuations of the eigenvector
amplitudes throughout the graph (see Eq. (\ref{hhssl})). The distribution of the LDOS attains a nonsingular, $\epsilon$-independent
limit as $\epsilon \rightarrow 0^{+}$, confirming the absence of localized eigenvectors in the high-connectivity limit \cite{Abou1973}.
The importance of degree fluctuations is more evident at the zero eigenvalue $\lambda=0$, where the
distribution of the LDOS
exhibits a power-law tail with exponent $\alpha+1$ (see Eq. (\ref{hkskd})).
In particular, the divergence of the mean value of the LDOS at $\lambda=0$ 
explains the singular
behaviour of $\rho(\lambda)$ for $\alpha \leq 1$.

It is interesting to compare our analytic expression for the distribution
of the LDOS at $\lambda=0$ with the analogous result for the extended phase of sparse regular random graphs with on-site disorder \cite{Mirlin1994,Mirlin1994A}. In the
latter class of models, the distribution of the LDOS decays exponentially fast and all its moments are finite, whereas in the present
model the $q$-th moment diverges for $\alpha \leq q$.
In particular, the second moment of the LDOS for regular random graphs
only diverges as the critical point
for the Anderson transition is approached from the delocalized phase by increasing the strength of the diagonal
disorder \cite{Evers2008,Tikhonov2019A,Tikhonov2019}.
In an analogous way, the second moment
of the LDOS in the present model is finite for $\alpha > 2$ and it diverges for
$\alpha \leq 2$, which seems to suggest
that highly-connected random graphs with strongly fluctuating degrees lie in a critical regime \cite{Mirlin2000,Evers2008}.
Besides constituting an interesting benchmark to study how degree heterogeneities affect the spectral
properties of networks, our analytic findings open the possibility
to investigate how the interplay between on-site disorder and fluctuations in the network
topology modify the Anderson localization transition.

Overall, our results uncover an interesting high-connectivity regime in which the resolvent
equations admit exact and nontrivial solutions that incorporate heterogeneous features of the network topology.
Thus, it would
be interesting to generalize the techniques developed in this work to solve the resolvent equations for
the adjacency matrix of directed random graphs \cite{Metz2019,Baron2022} and networks with loops \cite{Metz2011}, as well
as the analogous equations for the Laplacian matrix on graphs \cite{Wlod2006}.
Work along these lines is under way.

\ack

J.D.S. acknowledges a fellowship from CNPq/Brazil. F.L.M. thanks London Mathematical Laboratory and CNPq/Brazil for financial support.

\appendix{}
\section{Calculation based on characteristic functions}
\label{App_A}

In this appendix we present a more formal derivation of Eqs. (\ref{Ws}) for the probability
density $\mathcal{W}_z(s)$, from which all subsequent results for the spectral and localization properties follow.
By inspecting Eqs. (\ref{pdfresolven}) and (\ref{S}), we note that $\mathcal{W}_z(s)$ can be written as
\begin{equation}
  \label{W1}
  \fl
\mathcal{W}_z(s) = \sum_{k=0}^\infty p_k  \int_{\mathbb{H}^+} \left[\prod_{\ell=1}^{k}d g_\ell \mathcal{Q}_{z}(g_\ell)  \right] \int_{\mathbb{R}}
\left[ \prod_{\ell=1}^{k} d {J_{\ell}} p_J(J_\ell) \right]
\delta\Bigg ({s - \sum_{\ell=1}^{k}J_{\ell}^2 g_\ell}\Bigg),
\end{equation}
with $d g = d {\rm Re} g \, d {\rm Im} g$. 
In a similar fashion, one can introduce the distribution $\mathrm{W}_{z}(s)$ associated to $\mathcal{Q}_{z}(g)$. 
The average of an arbitrary function $f(G)$ of the cavity resolvent $G$ distributed according to $\mathcal{Q}_{z}(g)$,
\begin{equation}
\label{average_F}
    \langle f(G) \rangle  = \int_{\mathbb{H^+}} dg \mathcal{Q}_{z}(g) f(g),
\end{equation}
is recast in the form
\begin{equation}
\label{average_F11}
    \langle f(G) \rangle  = \int_{\mathbb{H^+}} ds \mathrm{W}_{z}(s) f \left( \frac{1}{z-s} \right) \quad d s = d {\rm Re} s \, d {\rm Im} s,
\end{equation}
where the expression for $\mathrm{W}_{z}(s)$ is inferred from Eq. (\ref{pdfresolvencav_Ninf})
\begin{equation}
  \label{Wcav1}
  \fl
\mathrm{W}_{z}(s) = \sum_{k=1}^\infty \frac{k}{c}p_k \int_{\mathbb{H}^+} \left[\prod_{\ell=1}^{k-1}d g_\ell \mathcal{Q}_{z}(g_\ell)  \right] \int_{\mathbb{R}}
\left[ \prod_{\ell=1}^{k-1} d {J_{\ell}} p_J(J_\ell) \right] \delta\Bigg ({s - \sum_{\ell=1}^{k-1}J_{\ell}^2 g_\ell}\Bigg).
\end{equation}
The quantity $\mathrm{W}_{z}(s)$ is the probability density of the random variable defined in Eq. (\ref{S}) with the replacement $k \rightarrow k-1$.
In particular, it follows from Eq. (\ref{average_F}) that the average resolvent $\langle G \rangle$ on the cavity graph is given by
\begin{eqnarray}
\label{A_G}
\langle G \rangle = \int_{\mathbb{H}^+} ds \frac{\mathrm{W}_{z}(s)}{z-s}.
\end{eqnarray}
The distributions $\mathcal{W}_{z}(s)$ and $\mathrm{W}_{z}(s)$ fully determine the spectral properties of the adjacency matrix. 

Our aim is to calculate the joint distributions $\mathcal{W}_{z}(s)$ and $\mathrm{W}_{z}(s)$ for $c \rightarrow \infty$. Given that $\mathcal{W}_{z}(s)$ and $\mathrm{W}_{z}(s)$
are distributions of sums of independent and identically distributed random variables, it is natural
to work with the characteristic functions of such distributions. Let $\mathcal{V}(u,v)$ and $\mathrm{V}(u,v)$ be the characteristic functions
of, respectively, $\mathcal{W}_{z}(s)$ and $\mathrm{W}_{z}(s)$, defined as
\begin{eqnarray}
\mathcal{V}(u,v) &=  \int_{\mathbb{H}^+} ds \mathcal{W}_{z}(s) \exp{\left(- i u {\rm Re} s -  i v {\rm Im} s   \right)}, \\
\mathrm{V}(u,v) &=  \int_{\mathbb{H}^+} ds \mathrm{W}_{z}(s) \exp{\left(- i u {\rm Re} s -  i v {\rm Im} s   \right)}.
\end{eqnarray}  
Inserting Eqs. (\ref{W1}) and (\ref{Wcav1}) in the above expressions, we obtain
\begin{eqnarray}
\label{phi}
    &\mathcal{V}(u,v) = \sum_{k=0}^\infty p_k \exp \big[k\mathcal{S}_c(u,v)\big],
\\
\label{phicav}    
    &\mathrm{V}(u,v) = \sum_{k=1}^\infty \frac{k}{c}p_k \exp \big[(k-1)\mathcal{S}_c(u,v)\big],
\end{eqnarray}
with
\begin{equation}
  \label{Sc}
  \fl
\mathcal{S}_c(u,v) = \ln \Bigg[\int_{\mathbb{H}^+}  d g \mathcal{Q}_{z}(g) \int_{-\infty}^{+\infty}d J p_J(J) \exp\Big(- i u  J^2 {\rm Re} g - i v  J^2 {\rm Im} g   \Big) \Bigg].
\end{equation}
Since the second moment of the coupling strengths is of $\mathcal{O}(1/c)$, the leading term of the above equation for $c \gg 1$ is given by
\begin{equation}
\mathcal{S}_c(u,v) = - i u  \frac{J_{1}^2}{c} {\rm Re} \langle G \rangle - i v  \frac{J_{1}^2}{c} {\rm Im} \langle G \rangle ,
\end{equation}  
where we assumed that $\langle G \rangle$ attains a well-defined limit for $c \to \infty$. The substitution of the above
expression for $\mathcal{S}_c(u,v)$ in Eqs. (\ref{phi}) and (\ref{phicav}) leads to the following equations for $c \rightarrow \infty$    
\begin{eqnarray}
\mathcal{V}(u,v) = \int_{0}^{\infty} d \kappa \, \nu(\kappa) \exp{\left(  - i u \kappa  J_{1}^2 {\rm Re} \langle G \rangle - i v \kappa  J_{1}^2 {\rm Im} \langle G \rangle    \right)}, \\
\mathrm{V}(u,v) = \int_{0}^{\infty} d \kappa \, \kappa \, \nu(\kappa) \exp{\left(  - i u \kappa  J_{1}^2 {\rm Re} \langle G \rangle - i v \kappa  J_{1}^2 {\rm Im} \langle G \rangle    \right)},
\end{eqnarray}  
where the probability distribution $\nu(\kappa)$ of the rescaled degrees is defined in Eq. (\ref{nu_kappa_def}). Performing
the inverse Fourier transform of $\mathcal{V}(u,v)$ and $\mathrm{V}(u,v)$, we get
\begin{eqnarray}
\label{W_nu}
&\mathcal{W}_{z}(s) = \int_{0}^{\infty} d\kappa\,\nu(\kappa) \delta(s - \kappa J_1^2 \langle G\rangle),\\
\label{Wcav_nu}
&\mathrm{W}_{z}(s) = \int_{0}^{\infty} d\kappa\,\kappa\,\nu(\kappa) \delta(s - \kappa J_1^2 \langle G\rangle).
\end{eqnarray}
Equation (\ref{W_nu}) means that the complex random variable $S$, distributed according to $\mathcal{W}_{z}(s)$, is equal in distribution to the
random variable $\kappa J_1^2 \langle G\rangle$. Thus, given  $\nu(\kappa)$, Eq. (\ref{Ws})  follows
by making a change of variables. The self-consistent equation for $\langle G \rangle$, Eq. (\ref{ffop}), is readily obtained
by inserting Eq. (\ref{Wcav_nu}) in Eq. (\ref{A_G}). This completes the calculation of $\mathcal{W}_{z}(s)$.

\section*{References}
\bibliography{bibliography}

\end{document}